\documentclass[aps,
groupedaddress,showpacs,nofootinbib,prl,twocolumn]{revtex4}
\usepackage{amssymb,amsmath}
\usepackage{graphics}
\usepackage[english]{babel}
\usepackage{graphicx}
\usepackage{dcolumn}
\usepackage{bm}

\usepackage{ulem}
\usepackage{cancel}
\usepackage{color}
\def\rr#1{\textcolor{red}{#1}}

\def\mn#1{\marginpar[\tiny{\rr{#1}}]{\tiny{\rr{#1}}}}

\def\rr#1{}

\usepackage{hyperref}

\def\comment#1{}

\newcommand{\be}{\begin{equation}}\newcommand{\ee}{\end{equation}}
\newcommand{\bea}{\begin{eqnarray}}\newcommand{\eea}{\end{eqnarray}}
\newcommand{\beaa}{\begin{eqnarray}}\newcommand{\eeaa}{\end{eqnarray}}
\newcommand{\ba}{\begin{array}}\newcommand{\ea}{\end{array}}
\newcommand{\bit}{\begin{itemize}}\newcommand{\eit}{\end{itemize}}
\newcommand{\ben}{\begin{enumerate}}\newcommand{\een}{\end{enumerate}}

 \newcommand{\sfrac}[2]{\raisebox{0.095ex}{\scriptsize${\frac{#1}{#2}}$}}

									\def\be{\begin{equation}}
\def\ee{\end{equation}}				\def\bea{\begin{eqnarray}}				\def\eea{\end{eqnarray}}
\def\bear{\begin{array}}				\def\eear{\end{array}}					
						\def\pa{\partial}							\def\pam{\partial^{\mu}}
				\def\pamm{\partial_{\mu}}					    
												
													\def\m{\mu}
\def\n{\nu}							\def\r{\rho}								\def\s{\sigma}
						\def\t{\tau}								\def\y{\psi}
														\def\p{\pi}
\def\a{\alpha}												\def\b{\beta}
\def\d{\delta}							\def\l{\lambda}												
						\def\G{\Gamma}						\def\D{\Delta}
						\def\W{\Omega}						\def\Y{\Psi}
												
\def\nn{\nonumber}											
												
												\def\x5{x^{5}}

													\def\fr{\frac}

\begin{document}

\title{Quantum Maupertuis Principle
}

\author{Antonia Karamatskou$^{(a)}$ and Hagen Kleinert$^{(a,b)}$}
\email{a.karamatskou@googlemail.com, \\ h.k@fu-berlin.de}


\affiliation{$^{(a)}$Institut f{\"u}r Theoretische Physik, Freie Universit\"at Berlin, 14195 Berlin, Germany}
\affiliation{$^{(b)}$ICRANeT Piazzale della Repubblica, 10 -65122, Pescara, Italy}


\vspace{2mm}

\begin{abstract}

According to the Maupertuis principle, the 
movement 
of  a classical particle in an external potential $V(x)$ can be understood
as the movement in a curved space with the metric 
$g_{\mu\nu}(x)=2M[V(x)-E]\delta_{\mu\nu}$. We show 
that the principle 
can be extended to the quantum regime, i.e.,
we show that the wave function of the particle
follows a 
Schr\"odinger equation 
in curved space
where the kinetic operator is formed with 
the {\it Weyl--invariant Laplace-Beltrami} operator.
As an application, we use DeWitt's recursive semiclassical expansion
of the time-evolution operator in curved space
to calculate 
the semiclassical expansion of the particle density
$\rho(x;E)=\langle x|\delta(E-\hat H)|x\rangle$.
 
\end{abstract}


\maketitle


The famous principle discovered in 1741 by Pierre Louis Maupertuis
and refined by Hamilton and Jacobi
laid the foundation to the geometric formulation of Newton's laws, and
was an important stimulus for Einstein's 
general theory of relativity.
In this note we want to point out that 
this geometric view of classical physics remains also valid
in the 
quantum regime, i.e.,
the quantum mechanics of a particle in a potential $V(x)$
may be described alternatively by
a Schr\"odinger equation 
in curved space with
 the Maupertuis metric 
\begin{eqnarray}
g_{\mu\nu}(x)\equiv 2M[V(x)-E]\delta_{\mu\nu}.
\label{@MM}\end{eqnarray}
The Hamiltonian of this Schr\"odinger equation
contains the 
 {\it Weyl-invariant\/}
(conformally-invariant) 
version  
\begin{align}
\D_{\rm W}=\D-\fr{1}{4}\fr{D-2}{D-1}R. \label{conflapl}
\end{align}
of the 
 Laplace-Beltrami operator 
\begin{align}
\D&=g^{-1/2}\pamm g^{1/2}g^{\m\n}\pa_\n.\label{LaplBeltr}
\end{align}
Our result supplies us with an answer to an old, very fundamental problem 
left open by Einstein's classical 
equivalence principle. That principle states that 
the classical laws of motion 
of a point particle 
can be derived from a coordinate
transformation in spacetime 
whose inertial forces  
simulate the gravitational 
forces at the position of the particle.
Since a quantum particle is always an extended
object described by a wave packet, 
there can be a correction term  $\xi R$ proportional to the curvature
scalar $R$,  whose size $\xi$ is undetermined by 
the classical 
equivalence principle. The Quantum Maupertuis
Principle
fixes the size of the $R$-term.

{\bf 1.}
Consider the  
{\it eikonal} of an arbitrary
trajectory of a point particle 
moving in Euclidean space  
with a potential $V(x)$ which
is defined as
\begin{eqnarray}
S(E)\equiv \int \sqrt{g_{\mu\nu}^E(x)dx^\mu
dx^\nu},
\label{@SS}\end{eqnarray}
with the Maupertuis metric 
(\ref{@MM}).
The integral (\ref{@SS})
is a functional of the trajectory which may be parameterized 
with the help of an arbitrary variable
$\lambda$ as $x^\mu(\lambda)$,
and rewritten as 
\begin{eqnarray}
S(E)\equiv \int d \lambda\sqrt{g_{\mu\nu}(x(\lambda))\dot x^\mu(\lambda)
\dot x^\nu(\lambda)}\equiv l.
\end{eqnarray}
The right-hand side coincides
with 
the  
invariant length of the trajectory.

According
to 
Maupertuis, 
the eikonal 
 $S(E)$ is extremal for
the classical trajectory, i.e., the classical orbit is {\it geodetic}.
If $\lambda$ is chosen to coincide with 
the invariant length $l$,
the extremization 
produces 
the geodetic differential equation
 \begin{align}
  \fr{d^2 x^\d}{dl^2}+\G^{\ \ \
    \d}_{\a\b}\fr{dx^\a}{dl}\fr{dx^\b}{dl}=0,
\label{EOM}
  \end{align}
where
$ {\G}_{\m\n}^{\ \ \l}$ are the Christoffel symbols
\begin{align}
 {\G}_{\m\n}^{\ \ \l}&=\fr{1}{2}{g}^{\l\s}\left(\pa_\m{g}_{\s\n}
 +\pa_\n{g}_{\m\s}-\pa_{\s}{g}_{\m\n} \right).\label{CHS}
\end{align}
Inserting 
the metric
(\ref{@MM}), we see that Eq.~(\ref{EOM}) is fulfilled if the trajectory 
follows the Newton equation $
x''^\m=-
\pam V.$

The Maupertuis metric 
(\ref{@MM})
differs from the flat Euclidean
metric $\bar g_{\mu\nu}\equiv \delta_{\mu\nu}$ only by a conformal factor
\begin{eqnarray}
\Omega^2(x)\equiv 2M[V(x)-E], 
\label{@OME}\end{eqnarray}
it is therefore called {\it conformally flat}.
The geometric properties 
of this space can be calculated
directly as functions of $
\Omega(x)$. We observe that under 
the {\it Weyl transformation}  $\bar g_{\mu\nu}(x) \rightarrow
g_{\mu\nu}=\Omega^2(x)\bar g_{\mu\nu}(x)$, the symbols
(\ref{CHS}) change like
\begin{align}
 \G_{\m\n}^{\ \ \l} &=\bar{\G}_{\m\n}^{\ \ \l}+\W^{-1}\left( \d^\l_\n \pa_\m \W +\d^\l_\m\pa_\n \W-\bar{g}^{\s\l}\bar{g}_{\m\n} \pa_\s \W\right). 
\end{align}
Because of this, the Riemann tensor defined by the covariant curl \cite{rem}%
\begin{align}
R_{\m\n\l}^{\ \ \ \ \s}&=\pa_\m \G_{\n\l}^{\ \ \s}-\pa_\n \G_{\m\l}^{\ \ \s}
-\G_{\m\l}^{\ \ \t}\G_{\n\t}^{\ \ \s}+\G_{\n\l}^{\ \ \t}\G_{\m\t}^{\ \ \s}\label{RIEM}
\end{align}
is related to 
${\bar R}_{\m\n\l}{}^\s$ by
\begin{align}
& R_{\m\n\l}^{\ \ \ \ \s}\!=\!\bar{R}_{\m\n\l}^{\ \ \ \s}\nn\\
 &\!+\!\left(2\bar{g}_{\l[\n}\d_{\m]\b}\bar{g}^{\s\a}\!-\!2\d^\s_{[\n}\d_{\m]\a}\d_{\l\b}
 +\d^\s_{[\n}\bar{g}_{\m]\l}\bar{g}^{\a\b} \right)\!\fr{\left(\pa_\a\W\right)\left(\pa_\b\W\right)}{\W^2}\nn\\
 &\!+\!\left( \d^\s_{[\n}\d_{\m]\a}\d_{\l\b}+\bar{g}^{\s\a}\bar{g}_{\l[\m}\d_{\n]\b}\right)\fr{\bar\nabla_\a \bar\nabla_\b\W}{\W},
\end{align}
where $\bar\nabla_ \mu v_\nu\equiv \pa_{\m}v_\n-  \bar{\G}_{\m\n}^{\ \ \l}v_\l$ 
stands for the covariant derivative, and $\bar{g}_{\l[\n}\d_{\m]\b}\bar{g}^{\s\a}$ 
is defined as the antisymmetrized expression 
$ 
\bar{g}_{\l[\n}\d_{\m]\b}\bar{g}^{\s\a}\equiv \bar{g}_{\l\n}\d_{\m\b}g^{\s\a}-\bar{g}_{\l\m}\d_{\n\b}\bar{g}^{\s\a}.$
The Ricci scalar $R\equiv g^{\nu\lambda}R_{\mu\nu\lambda}{}^\lambda$
is obtained from (\ref{RIEM}) as
\begin{align}
R&=\fr{\bar{R}}{\W^2}-2(D-1)\bar{g}^{\a\b}\fr{\bar\nabla_\a \bar\nabla_\b\W}{\W^3}\nn\\
&-(D-1)(D-4)\bar{g}^{\a\b}\fr{\left(\pa_\a\W\right)\left(\pa_\b\W\right)}{\W^4}.\label{Rconf}
\end{align}
Inserting
$\Omega(x)$ from 
(\ref{@OME}), 
this becomes
\begin{align}
R=\fr{1-D}{4}\left[ \fr{2\pam\pamm V}{M(E-V)^2}
+\fr{(D-6)\pam V\pamm V}{2M(E-V)^3} \right].\label{curvaturescalar}
\end{align}


{\bf 2.}
\def\nablab{\mbox{\boldmath$\nabla$}}
Consider now the quantum mechanics of the point particle
of energy $E$
in the potential $V(x)$. It is described by the
Schr\"odinger equation 
    \begin{equation}
(\hat H-E)\y(x)\equiv\left(  \fr{\hat {\bf p}^2}{2M}+V(x)-E\right)\y(x)=0,
  \label{SCHREQ} \end{equation}
where $\hat {\bf p}\equiv -i\hbar {\nablab}$.
Using the metric (\ref{@MM}),
this can be rewritten as 
$\left[ \Omega^{-2}(x)\hat {\bf p}^2+1\right]\y(x)=0$, or as
\begin{eqnarray}
\left[\hbar ^2\Delta_{\rm W} -1\right]\y(x)=0,
 \label{QMP} \end{eqnarray}
where $\Delta_{\rm W} \equiv\Omega^{-2}(x)\sum_{\mu}\partial_{x^\mu}^2$.
It is easy to verify that this is equal to the Weyl-invariant 
combination (\ref{conflapl})
of the Laplace-Beltrami operator (\ref{LaplBeltr})
and $R$.

Equation (\ref{QMP})
is a simple but very fundamental result.
The Maupertuis metric (\ref{@MM})
governs not only the classical 
motion, but also the quantum mechanics, provided that 
the 
 Laplace-Beltrami operator
is extended to the Weyl-invariant 
form
(\ref{conflapl}).

{\bf 3.}
The advantage of the curved-space 
reformulation
(\ref{QMP}) of the 
Schr\"odinger equation
(\ref{SCHREQ}) is that,
in curved space, the particle is without a potential.\mn{{\rr without this remark about free particle movement...}}
It is a {\it free  particle} moving through the Maupertuis metric (\ref{@MM}).
For such movements,
there exist
well-developed methods of calculating  
quantum properties 
pioneered by Bryce DeWitt
\cite{dew,leshou}.
In particular,
DeWitt has given a semiclassical expansion 
of the matrix elements of the resolvent operator 
\begin{eqnarray}
\langle x|\hat{\cal  R}|x'\rangle \equiv \langle x|\frac{i\hbar
}{{\cal E}-\hat{\cal H}}|x'\rangle,
\label{@res}\end{eqnarray}
where 
$\hat{\cal H}$ is a curved-space
translation operator in some
{\it  pseudotime}
parameter $\tau$,
 and ${\cal E}$ is the associated pseudoenergy.
The pseudotime $\tau$ is commonly called  
{\it Schwinger time} or {\it fifth time}.
The Green function 
$
G(x,x')=\langle x| \hat R| x' \rangle
$
can be written as
an integral
\begin{eqnarray}
G(x,x')= \int_0^\infty d \tau
\langle 
x,\tau|x',0\rangle
\label{intequ}
\end{eqnarray}
over the 
 pseudotime displacement 
amplitude
\begin{eqnarray}
\langle 
x,\tau|x',0\rangle
=\langle x  | e^{-i(\hat{\cal H}-{\cal E})\tau/\hbar}|x'\rangle.
\end{eqnarray}
This
amplitude
satisfies the 
 Schr\"odinger equation
\begin{align}
i\hbar\partial_\tau\langle x,\tau\mid x',0\rangle= \hat {\cal H}\,\langle
x,\t| x',0\rangle
\label{schroedequampl}
\end{align}
with the boundary condition in $D$ dimensions
\begin{align}
\langle x,0\mid x',0\rangle=\d^{(D)}(x-x').\label{boundcond}
\end{align}
The Lagrangian treated by 
DeWitt is
  \begin{align}
{\cal   L}=\fr{1}{2}g_{\m\n}(x)\dot{x}^\m\dot{x}^\n.
  \end{align}
This has the pseudotime Hamiltonian
${\cal H}=\sfrac12 g^{\m\n}(x)p_\mu p_\n\equiv
\sfrac12 p^\mu p_\n
$,
where
$ g^{\m\n}(x)$
is the inverse of
the metric
$ g_{\m\n}(x)$, and the action
  \begin{align}
{\cal A}(x,x';\tau-\tau')=\int^{x',\t'}_{x,\t}d\tau\,{\cal L} =\fr{\s(x,x')}{\t-\t'}.
  \end{align}
 where $\s(x,x')\approx \sfrac12 g_{\mu\nu}(x)(x-x')^\mu(x-x')^\nu+\dots$
 is the geodetic interval. 
The action depends on the pseudotime only via this ratio. 
This is a consequence of the ``free motion'' in the metric $g_{\mu\nu}(x)$.  

From the Hamilton-Jacobi equations it follows
that
    \begin{align}
   \fr{\pa {\cal A}}{\pa x^\m}&= p_\m=\fr{\s_{\m}}{(\t-\t')},\\
    -\fr{\pa  {\cal A}}{\pa \t}&=\fr{\s(x,x')}{(\t-\t')^2}={\cal H}=\fr{1}{2}p_\m p^\m.
    \end{align}
DeWitt gave the solution of the Schr\"odinger equation (\ref{schroedequampl})
as a power series in $\tau$
for the Hamiltonian 
\begin{eqnarray}
\hat{\cal H}=\sfrac12\left(-\Delta+\xi R +m^2\right),
\label{dewham}\end{eqnarray}
with an arbitrary parameter $\xi$.
For small  $\tau$  and $x$ close to $x'$,
the solution
is simply
\begin{align}
\langle x,\tau\mid x',\tau'\rangle\approx\frac{ D_{\rm MV}^{1/2}(x,x')}{ (2\p i\hbar s)^{D/2}}e^{i\s(x,x')/s\hbar},\label{seriesexp0}
\end{align}
 where 
$s\equiv \t-\t'$ and
$D_{\rm MV}\equiv
\det [-\partial_\mu \partial ' _\nu
\sigma(x,x')
]
$  is the Morette-van
Vleck determinant \cite{morse, rem1}.
For arbitrary $s$, the result is (\ref{seriesexp0})
\begin{align}
\!\!\!\langle x,\tau|x',\tau'\rangle=
\frac{ D_{\rm MV}^{1/2}(x,x')}{ (2\p i\hbar s)^{D/2}}
e^{i\s(x,x')/s\hbar}\sum_{n=0}^{\infty}a_n(is/2\hbar)^n,
\label{seriesexp}
\end{align}
where 
$
D^{1/2}_{\rm MV}(x,x') 
\equiv
g^{1/4}(x)
\D^{1/2}_{\rm MV}(x,x') 
g^{1/4}(x')
$ and 
$\D_{\rm MV}(x,x')$
has
the endpoint expansion (i.e., the derivatives are evaluated at the endpoint $x$)
\begin{align}
&\D_{\rm MV} ^{1/2}  \!=\! 1\! +\! \fr{1}{12}R_{\m \n } \s ^\m  \s ^\n
\!\! -\! \fr{1}{24}R_{\m \n;\r } \s ^\m  \s ^\n  \s ^\r     \\  &
\! +\hspace{-1pt}\!\left(\!\frac{1}{288}\!R_{\m \n } R_{\r \t }  \!+ \!\fr{1}{360}\!\hspace{-1pt}R^{\a\
    \!\!\b} _{\ \! \m\ \hspace{-1pt} \n}  R_{\a \r \b \t } \! +\!
  \fr{1}{80}\hspace{-1pt}\!
R_{\m \n;\r \t  }\!\! \right)\!\hspace{-1pt}\s ^\m \! \s ^\n\!  \s
^\r\!  \s ^\t\! \!  
+\!\ldots \hspace{-1pt}.\nn \label{detexp}
\end{align}
DeWitt  allowed for the presence of an extra term $\xi R$ in addition
to the Laplace-Beltrami operator $\Delta$ on the right-hand side 
of (\ref{schroedequampl}).
Then he derived a recursion relation for the 
expansion coefficients \cite{leshou}
\begin{align}
\s_\m(a_0)_;^{\ \m}&\!=\!0\\\!\!\!\!\!\!
(n\!+\!1)a_{n+1}+\!\s_\m(\!a_{n+1}\!)_;^{\ \m}&\!=\!\D_{\rm MV}^{-1/2}\!\left(\! \D_{\rm MV}^{1/2}a_n\! \right)\!_{;\m}^{\ \ \m}\!-\!\xi R a_n ,\label{recrel}
\end{align}
whose lowest terms are
\begin{align}
a_1 
&=\left( \fr{1}{6}-\xi\right)R\label{coeffa1}
\\
a_2 
&=\fr{1}{6}\left( \fr{1}{5}-\xi \right)R_{;\m}^{\ \ \m}+\fr{1}{2}\left( \fr{1}{6}-\xi \right)^2R^2\nn\\
		&-\fr{1}{180}R_{\m\n}R^{\m\n}+\fr{1}{180}R_{\m\n\r\s}R^{\m\n\r\s}. \label{coeffa2}
\end{align}

{\bf 4.}
We now come to 
the announced application 
of the quantum Maupertuis principle by
calculating the particle density
of the Schr\"odinger equation (\ref{SCHREQ})
\begin{eqnarray}
\rho(x;E)\equiv\langle x|
\delta(E-\hat H)|x\rangle=\fr{1}{2\p \hbar}{\rm disc}\left(\fr{i\hbar}{E-E_n}\right).
\label{@PD}\end{eqnarray}
A simple algebra shows that
\begin{eqnarray}
\langle x|\hat R|x'\rangle \!=\!\fr{1}{2}
\langle x|\hat {\cal R}|x'\rangle
 [V(x')\!-\!E]^{-1}.
\label{extra}
\end{eqnarray}
Now we insert the  DeWitt 
expansion
(\ref{seriesexp}) 
which reduces for  $x=x'$
to 
\begin{eqnarray}\!\!
\langle x|\hat {\cal R}|x\rangle\!
=\!\frac{g^{1/2}(x)}{(2\pi i\hbar)^{D/2}}\!\sum _{n=0}^\infty\! a_n\,(\!-\partial_{m^2}\!)^n\!\! \int_0^\infty
\!\frac{ds\, e^{-im^2 s/2\hbar}}{s^{D/2}},
\label{@exp}\end{eqnarray}
where 
the integral is simply
$\Gamma(1-D/2)(m^2) ^{D/2-1}$, so that
the sum on the right-hand side becomes
\begin{align}
\sum _{n=0}^\infty a_n \,\Gamma(n+1-D/2)(m^2) ^{D/2-(n+1)} .\end{align}
To be used in
in Eq.~(\ref{extra})
we must take Eq.~(\ref{@exp})
for 
$\xi=(D-2)/4(D-1)$ and $m^2=1$
and evaluate 
 $a_n$
with 
curvature terms of
the Maupertuis metric (\ref{@MM}), where
\begin{align}
&\!\!\!
\langle x|\hat {\cal R}|x\rangle=
\left(\fr{M}{2\p\hbar^2}\right)^{D/2}\left\{ \G(1-D/2)(V-E)^{D/2}\right.\nn\\
&\!\!\!\left.-\fr{\hbar^2}{12 M}\G(3-D/2)\pamm \pam V(V-E)^{D/2-2} \right.\nn\\
&\!\!\!\left.+\fr{\hbar^2}{24M}\G(4-D/2)\right.\left. \pamm V\pam
  V(V-E)^{D/2-3}+\ldots \right\} 
.\label{geomden}
\end{align}
The result is valid 
for $V(x)>E$ where the metric is positive.
For $E>V(x)$ se use the property
 $
V-E=e^{\mp i\p}(E-V)
$ to find the discontinuity across the cuts.
Remembering the extra factor $(V-E)^{-1}$ in (\ref{extra})
we obtain from the  DeWitt expansion 
the particle density
$\r_{_{\rm{DW}}}(x;E)\equiv\langle x| \delta({\cal E}-\hat {\cal
H})|x\rangle $ as
\begin{align}
&\r_{_{\rm{DW}}}(x;E)=\fr{1}{\p}\left(\fr{M}{2\p \hbar^2}\right)^{D/2}
\sin\left(\fr{\p D}{2}\right)\nn\\ 
&\times\left[\G(1-D/2)(E-V)^{D/2-1}\right.\nn\\
&\left.-\fr{\hbar^2}{12M}\G(3-D/2)(E-V)^{D/2-3}\pamm \pam V\right.\label{fordens}\\
&\left.-\fr{\hbar^2}{24M}\right. \left.\G(4-D/2)(E-V)^{D/2-4}\pamm V\pam V+\ldots\right],\nn
\end{align}
Now we employ
the reflection formula
for Gamma functions
$
\G(1-z)\G(z)={\p}/{\sin(\p z)}
\label{reflform}
$ 
to find 
\begin{align}
& \r_{_{\rm{DW}}}(E;x)=\left(\fr{M}{2\p \hbar^2}\right)^{D/2}\left[\fr{1}{\G(D/2)}(E-V)^{D/2-1}\right.\nn\\
&\left.-\fr{\hbar^2}{12M}\fr{1}{\G(D/2-2)}(E-V)^{D/2-3}\pamm \pam V\right.\label{densityhk}\\
&\left.+\fr{\hbar^2}{24M}\right.\left.\fr{1}{\G(D/2-3)}(E-V)^{D/2-4}\pamm V\pam V+\ldots\right]\nn.
\end{align}
This agrees with the result 
obtained from  
the original Schr\"odinger equation 
(\ref{SCHREQ}) $E>V(x)$ \cite{rem1}.

{\bf 5.}
By virtue of the bilocal character of the DeWitt
techniques in curved space,
the expansion 
of $\langle x|\hat {\cal R}|x\rangle$ exists also 
for the off-diagonal matrix elements
$\langle x|\hat{\cal R}|x'\rangle$
which serves to
 find also
the  
off-diagonal particle density
$
\rho(x;E)\equiv\langle x|
\delta(E-\hat H)|x'\rangle$
beyond the result stated in the literature 
 \cite{rem2}.

{\bf 6.}
The extra $R$-term found above is not universal.
This can be seen by comparing the result with the quantum mechanics 
of another system in curved space:
the hydrogen atom in momentum space. It 
 obeys a Schr\"odinger equation 
\be
\left( {\bf p}^2+p_E^2\right) \Y(p)=\fr{2}{\hat r}\Y(p).
\label{MOMSP}\ee
Here
$\hat r$ is operator 
of the radial coordinate  
n the momentum representation,
and 
 $p_E^2=-2E$ 
(in
natural units with $\hbar= a_H=E_H=1$, where
 $a_H\equiv \alpha ^2 \hbar /m_e c\,$=\,Bohr radius
and $E_H=\alpha^2 m_ec^2\,$=\,Rydberg energy).
 By analogy
with the previous approach 
we rewrite (\ref{MOMSP})
as
\begin{eqnarray}
\{\sfrac14[\hat r({\bf p}^2+p_E^2)]^2-1\}\Y(p)=0.
\label{@SEP}\end{eqnarray}
Reordering this we can 
bring the two operators  $\hat r$
side by side
to express  $\hat r^2$ as $\sum_{\mu=1}^D\partial^2_{p^\mu}$,
and 
(\ref{@SEP}) turns into 
the differential equation 
\begin{eqnarray}
\left(\fr{1}{2}\D_p-p_E^2
+1 \right)\Y=0.
\label{@xx}\end{eqnarray}
where 
$\Delta_p$ is now the Laplace-Beltrami operator 
in momentum space formed from the metric
\be
g_{ij}=\frac{2}{( {\bf p}^2+p_E^2)^2}\delta_{ij},\label{metrichatom}
\ee
which is again conformally flat.
The associated  curvature scalar is now
$R=2D(D-1)p_E^2$,
so that 
(\ref{@xx}) can be rewritten as
\begin{eqnarray}
\left(\fr{1}{2}\D_p-
\fr{R}{2D(D-1)}
+1 \right)\Y=0.
\label{@xx}\end{eqnarray}
Remarkably,  the coefficient of the $R$-term 
in this momentum space problem
does {\it not} correspond to the Weyl-invariant expression, 
where the subtracted $R$-term would have been $(D-2)R/8(D-1)=R/16$ for $D=3$.

{\bf 6.}
The result 
gives us the possibility
of studying the quantum mechanics of an arbitrary potential 
problem  
using the well-developed 
techniques of curved-space quantum mechanics \cite{leshou}.
Conversely, it permits us to understand questions
about the quantum mechanics in curved
space from the knowledge of Schr\"odinger theory in flat space.

~\\
Acknowledgement:
We are grateful to Axel Pelster
for discussions.


\end{document}